\documentclass[conference]{IEEEtran}
\IEEEoverridecommandlockouts
\usepackage{cite}
\usepackage{amsmath,amssymb,amsfonts}
\usepackage{algorithmic}
\usepackage[ruled,vlined]{algorithm2e}
\usepackage{subfig}
\usepackage{graphicx}
\usepackage{textcomp}
\usepackage{xcolor}
\usepackage{float}
\usepackage{multirow}
\usepackage{epstopdf}
\usepackage{mathtools}
\usepackage{url}

\begin{document}
\title{Keep It Simple: CNN Model Complexity Studies for Interference Classification Tasks}
\author{\IEEEauthorblockN{Taiwo Oyedare$^{\star}$, Vijay K. Shah$^{\dagger}$, Daniel J. Jakubisin$^{{\star}{\ddagger}}$, Jeffrey H. Reed$^{\star}$} 
$^{\star}$ Bradley Dept. of Electrical and Computer Engineering, Virginia Tech, Blacksburg, USA \\
$^{\dagger}$ Dept. of Cybersecurity Engineering, George Mason University, Fairfax, USA\\
$^{\ddagger}$ Virginia Tech National Security Institute, Blacksburg, VA, USA
\\
Email: toyedare@vt.edu, vshah22@gmu.edu, djj@vt.edu, reedjh@vt.edu
}

\maketitle

\begin{abstract}
The growing number of devices using the wireless spectrum makes it important to find ways to minimize interference and optimize the use of the spectrum. Deep learning models, such as convolutional neural networks (CNNs), have been widely utilized to identify, classify, or mitigate interference due to their ability to learn from the data directly. However, there have been limited research on the complexity of such deep learning models. The major focus of deep learning-based wireless classification literature has been on improving classification accuracy, often at the expense of model complexity. This may not be practical for many wireless devices, such as, internet of things (IoT) devices, which usually have very limited computational resources and cannot handle very complex models. Thus, it becomes important to account for model complexity when designing deep learning based models for interference classification. To address this, we conduct an analysis of CNN based wireless classification that explores the trade-off amongst dataset size, CNN model complexity, and classification accuracy under various levels of classification difficulty: namely, interference classification, heterogeneous transmitter classification, and homogeneous transmitter classification. Our study, based on three wireless datasets, shows that a simpler CNN model with fewer parameters can perform just as well as a more complex model, providing important insights into the use of CNNs in computationally constrained applications.  
\end{abstract}


\section{Introduction}
\label{sec:intro}

The proliferation of internet of things (IoT), 5G  devices, and other wireless technologies has led to an increase in the number of wireless devices that interfere with each other, either intentionally or unintentionally. Classifying interference is essential for ensuring good communication quality and adhering to spectrum sharing policies. The concept of wireless interference, where a signal of interest is disrupted by another signal sharing the same channel, has been studied extensively in the field of wireless communication \cite{qiu2007general,bruno2007interference,schmidt2017wireless}. This type of interference can significantly degrade the signal-to-noise-plus-interference ratio (SINR) and disrupt communication between a transmitter and receiver.

Traditionally, many interference classification techniques have relied on rule-based approaches that are not effective when multiple coexisting technologies are in use. As a result, alternative classification algorithms that use feature detection or extraction techniques, such as cyclostationary feature detection \cite{kim2007cyclostationary}, have been explored. However, these techniques require domain expertise and can result in a complicated solution that is oftentimes not scalable. In recent years, researchers have turned to deep learning techniques to reduce the need for domain expertise \cite{yu2020interference,kim2020classification,schmidt2017wireless}. Interference suppression applications have also widely used deep learning \cite{oyedare2022interference}. 

Convolutional neural networks (CNNs), a model-free deep learning approach, have been shown to be effective in various domains such as image classification and natural language processing. CNNs have been used for various classification tasks, including protocol/interference classification \cite{kim2020classification,yu2020interference,zhao2019indoor,schmidt2017wireless,grunau2018multi,sankhe2019oracle}, transmitter classification \cite{oyedare2019estimating}, and modulation classification \cite{o2016convolutional}. In using CNNs or other deep learning models for classification tasks in wireless communication applications, researchers need to ensure that they have access to high quality datasets and efficient models. 
While earlier works \cite{kim2020classification, grunau2018multi,yu2020interference} have focused on improving the classification performance of different deep learning models, the relationship between the size of a dataset, the complexity of the CNN models used, and the difficulty of classification is often overlooked or assumed. Even though many researchers would tune hyper-parameters of their CNN during the training process, an insight into other factors (such as filter size, number of nodes in the hidden layer, etc) that affect model complexity has not been typically investigated. In resource-constrained applications like the IoT, CNN models that are too complex may not be feasible for classification in real-world situations. This is because  IoT is marked by its limited processing power and storage capabilities, which can lead to challenges in terms of performance, security, privacy, and reliability \cite{atlam2018fog,pereira2020challenges}. It is essential to understand the processes that lead to the selection of hyper-parameters in relation to model complexity, the size of the dataset and the difficulty of classification.

A typical CNN architecture consists of a series of feed-forward layers that apply convolutional filters and pooling operations, followed by fully-connected layers that convert the 2D feature maps produced by the earlier layers into 1D vectors for classification \cite{ravi2016deep}. While CNNs do not require a separate feature extraction step before being used, they can be time-consuming and difficult to train from scratch because they require a large labeled dataset for building and training the model \cite{mohsen2018classification}. The complexity of deep learning models can be influenced by various factors, such as the number of layers, number of filters, size of the filters, and number of nodes in the hidden layer. Researchers in the field of deep learning often aim to improve the performance of their models by hyper-parameter tuning and other optimization techniques. In the literature, there has been a focus on improving classification performance through these methods. Although it is important to optimize deep learning models for performance, there has been limited attention in the wireless communication literature on thoroughly analyzing the factors that influence model complexity. This paper aims to fill this gap by studying the relationship between model complexity, dataset size, and classification difficulty in a thorough and empirical manner. To the best of our knowledge, this is one of the first studies to examine this relationship. 

Our contributions are as follows:
\begin{itemize}
\item We thoroughly analyze the complexity of three different CNN architectures (simple, medium, and complex) in relation to dataset size and classification difficulty.
\item We show, empirically, that the performance of a simple CNN model with fewer parameters is comparable to that of a more complex CNN model. This is important because resource-constrained devices, which have limited processing power and storage capabilities, can benefit from using simpler CNN models.
\end{itemize}

\section{Overview of Classification Tasks}
\begin{figure}[!tb]
    \centering
    \includegraphics[width=0.45\textwidth]{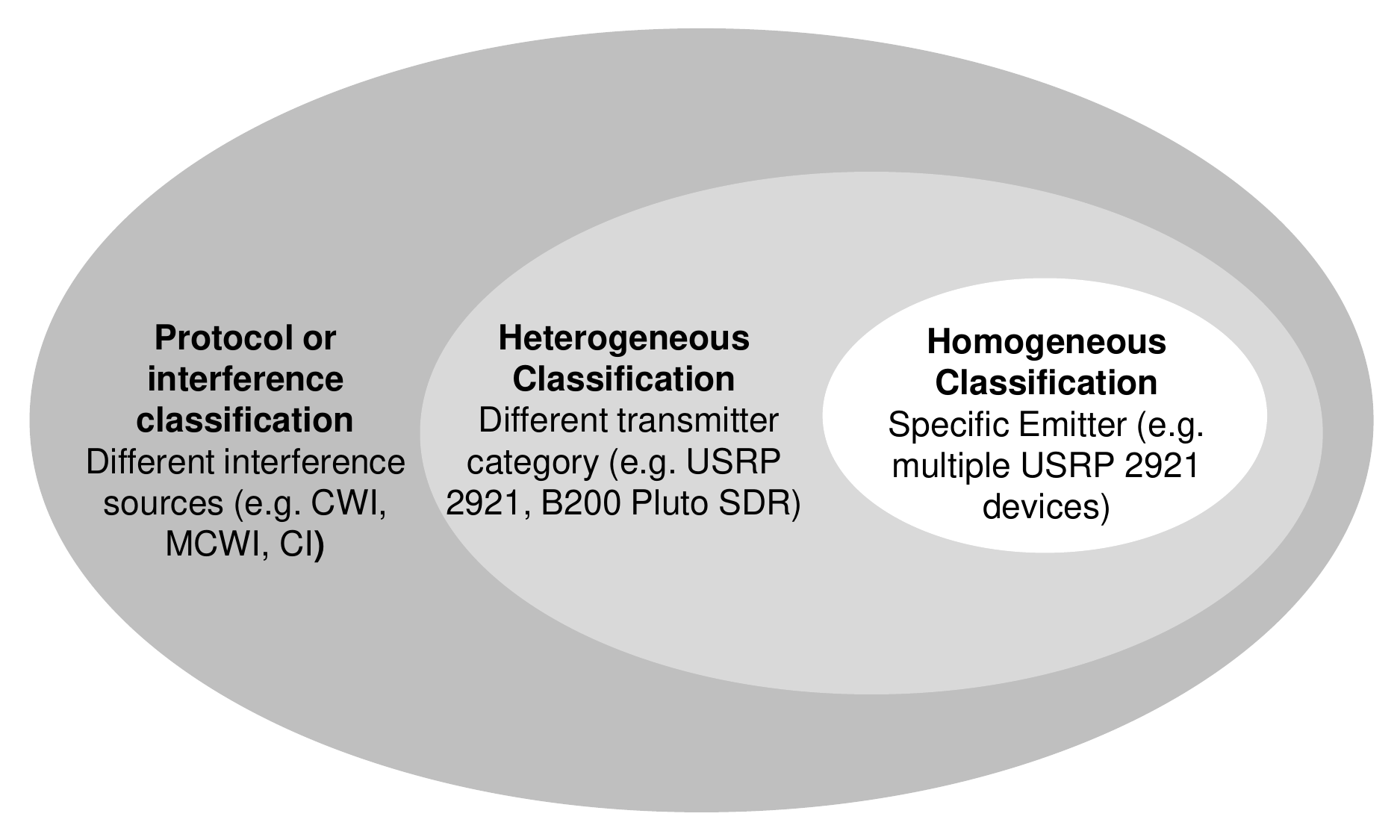}
    \caption{Overview of Interference Classification}
    \label{fig:overview}
\vspace{-0.2in}
\end{figure}

In this paper, our interference classification task is performed at levels of difficulty as shown in Fig. \ref{fig:overview}. At the interference or protocol level, a CNN can be used to classify different protocols or interference sources. At the heterogeneous level, a CNN can be used to classify different transmitter categories. Finally, at the homogeneous level, a CNN can be used to classify specific emitters (homogeneous) categories, such as transmitters from the same model or manufacturer. The outermost layer is the easiest classification while the innermost layer is the most difficult.

\subsection{Transmitter Categorization}
\label{class}
Most transmitters have features that are peculiar to each of them. For instance, when run at high power, the power amplifiers used in many wireless devices sometimes display non-linearities \cite{edalat2003effect}\cite{polak2011identifying}. These non-linearities can be used to group the transmitters into different categories. We briefly discuss the features of the transmitters used for our classification tasks.

\subsubsection{Category A (USRP 2921)}
These transmitters utilize more reliable linear power amplifiers and finer filters than other transmitter categories. 

\subsubsection{Category B (USRP B200)}
In comparison to those in category A, the components in the transmitters in this category are less reliable. They were designed with low-cost experimentation in mind. They employ a single chain of the $AD 9364$, which is frequently utilized to decrease hardware and software complexity.

\subsubsection{Category C (Adalm Pluto SDR)}
This category's transmitters are far less capable than the other two categories'. Considering their small size and low cost, the Pluto SDR is capable of a wide range of useful SDR applications. 

\subsection{Levels of Classification}
In this section we discuss the three levels of classification experiments carried out for the model complexity study. 

\subsubsection{Protocol or Interference Classification}
\label{protocol_class}
The protocol or interference level of classification is easier than the other two discussed in this section. This is because, there are enough distinguishing features at this level. For instance, in the radio frequency interference dataset used in this paper, there is a marked difference between the three types of jammers described . For instance, the MCWI, which combines the SoI and a two-tone CW is structurally different from the CI \cite{ujan2020efficient}. 

\subsubsection{Heterogeneous Classification}
One transmitter from each of the three categories listed above in Section \ref{class} is used in this study. While two of the three transmitters (USRP 2921 and USRP B200) were produced by the same company (National Instruments), all three are of distinct models. In level of difficulty, this classification is easier than homogeneous classification (discussed in Section \ref{homogenous_class}) but harder than the protocol or interference classification discussed in Section \ref{protocol_class} 

\subsubsection{Homogeneous Classification}
\label{homogenous_class}
In this classification task, we seek to distinguish transmitters within categories A, B, or C. The transmitters are identical (same manufacturer and model), making this the most challenging classification problem of all three levels. This is because the classification algorithm must identify slight variations in transmitters which have the same architecture and hardware components. The same OFDM waveform is sent by all of the devices, significantly complicating categorization.

\section{Implementation Details of the CNN}
\label{sec:CNN}
\subsection{CNN Parameters for Model Complexity}
\label{cnn_transmitter}
The CNN algorithm utilized was a modified \emph{Tensorflow} CNN model that was used to categorize handwritten digits from the MNIST dataset. We created three levels of complexity for our CNN models by varying the number of nodes in the hidden layer, the number of filters, and the size of the training dataset. These factors all contribute to the number of parameters in the model, and as a general rule, the complexity of a CNN algorithm increases with the number of parameters. As a CNN architecture becomes more complex, it is generally expected that the performance of the CNN algorithm will improve. However, one potential downside is that the algorithm may begin to overfit to the training data, resulting in a higher training accuracy compared to the test accuracy. 

Our network contains three convolutional layers, max pooling layers and a fully connected layer. A  $3\times 3$ filter  is applied to the input matrix by the convolutional layers. Convolution operations are carried out in the resulting sub-region to produce a single value in the respective output feature map. To integrate non-linearities into the model, the scale-invariant rectified linear unit (ReLU) activation function is applied to the feature map values.

The data collected by the convolutional layer is down-sampled using the pooling layer. We utilized the maximum pooling function of $2\times 2$. This indicates that the most important features of the signals are kept while others are deleted \cite{wang2016multiple}, which facilitates transmitter classification. Table \ref{parameter} lists specifics of the parameters for all three designs.

\begin{table}
\caption{CNN Parameters for Model Complexity}
\centering
\begin{tabular}{|p{2.5cm}|c|c|c|} 
\hline
\multicolumn{1}{|c|}{\textbf{Parameters}} & \textbf{Simple} & \textbf{Medium} & \textbf{Complex} \\ 
\hline
Nodes in the hidden layer & 0.26k & 1.04k & 8k \\ 
\hline
Total number~of parameters~for the whole~network & 6.1k & 48k & 276k \\ 
\hline
Filter size (No. of~Filters) & 3*3(16) & \multicolumn{2}{c|}{3*3(32)} \\ 
\hline
Batch size & \multicolumn{3}{c|}{16} \\ 
\hline
Number of layers & \multicolumn{3}{c|}{3} \\ 
\hline
Strides & \multicolumn{3}{c|}{2} \\ 
\hline
Number of classes & \multicolumn{3}{c|}{\begin{tabular}[c]{@{}c@{}}4 (for homogeneous devices)\\3 (for heterogeneous devices)\end{tabular}}  \\ 
\hline
Learning rate & \multicolumn{3}{c|}{0.0002} \\ 
\hline
Max pooling size & \multicolumn{3}{c|}{2 * 2} \\ 
\hline
Dropout Probability & \multicolumn{3}{c|}{50\%} \\ 
\hline
Input matrix size & \multicolumn{3}{c|}{38*100} \\ 
\hline
Activation function & \multicolumn{3}{c|}{ReLU} \\ 
\hline
Optimizer & \multicolumn{3}{c|}{Adam optimizer with cross entropy loss}\\ 
\hline
Loss & \multicolumn{3}{c|}{Cross entropy loss} \\ 
\hline
Training sizes & \multicolumn{3}{c|}{16,64,256,1024,4096,8192,16384} \\ 
\hline
Test sizes & \multicolumn{3}{c|}{2k} \\
\hline
\end{tabular}
\label{parameter}
\vspace{-0.15in}
\end{table}

\begin{table}[!htb]
\caption{CNN Parameters for Interference Classification}
\centering
\begin{tabular}{|c|c|}
\hline
\textbf{Parameters} & \textbf{RESNet} \\
\hline
Batch size & 64\\
\hline
Number of layers & 18 \\
\hline
Learning rate & 0.0001 \\
\hline
Maximum pooling dimension & $2\times 2$ \\
\hline
Activation function & ReLU \\
\hline
Training/test size & 80\%/20\% \\
\hline
Dropout Probability & 80\%  \\
\hline
\end{tabular}
\label{table:interf_class}
\vspace{-0.2in}
\end{table}

\begin{table*}[!hbt]
\caption{Comparison of Training Accuracy and Testing Accuracy for all Model Complexity and Device Category}
\centering
\begin{tabular}{ccccccccccccc}
\hline
\multicolumn{1}{l}{} & \multicolumn{6}{c}{\textbf{Homogeneous Classification (USRP 2921)}}  & \multicolumn{6}{c}{\textbf{Homogeneous Classification (USRP B200)}}\\ \hline
\multicolumn{1}{l}{}  & \multicolumn{2}{c}{\textbf{Simple CNN}}  & \multicolumn{2}{c}{\textbf{Medium CNN}} & \multicolumn{2}{c}{\textbf{Complex CNN}} & \multicolumn{2}{c}{\textbf{Simple CNN}} & \multicolumn{2}{c}{\textbf{Medium CNN}} & \multicolumn{2}{c}{\textbf{Complex CNN}} \\ \hline
\textbf{\begin{tabular}[c]{@{}c@{}}Dataset\\ Size\end{tabular}} & \textbf{\begin{tabular}[c]{@{}c@{}}Training\\ Acc\end{tabular}} & \textbf{\begin{tabular}[c]{@{}c@{}}Test\\ Acc\end{tabular}} & \textbf{\begin{tabular}[c]{@{}c@{}}Training\\ Acc\end{tabular}} & \textbf{\begin{tabular}[c]{@{}c@{}}Test\\ Acc\end{tabular}} & \textbf{\begin{tabular}[c]{@{}c@{}}Training\\ Acc\end{tabular}} & \textbf{\begin{tabular}[c]{@{}c@{}}Test\\ Acc\end{tabular}} & \textbf{\begin{tabular}[c]{@{}c@{}}Training\\ Acc\end{tabular}} & \textbf{\begin{tabular}[c]{@{}c@{}}Test\\ Acc\end{tabular}} & \textbf{\begin{tabular}[c]{@{}c@{}}Training\\ Acc\end{tabular}} & \textbf{\begin{tabular}[c]{@{}c@{}}Test\\ Acc\end{tabular}} & \textbf{\begin{tabular}[c]{@{}c@{}}Training\\ Acc\end{tabular}} & \textbf{\begin{tabular}[c]{@{}c@{}}Test\\ Acc\end{tabular}} \\ \hline
16 & 1.00 & 0.407 & 1.00 & 0.406 & 1.00 & 0.408 & 1.00 & 0.502 & 1.00 & 0.492 & 1.00 & 0.544 \\ \hline
64 & 0.996 & 0.545 & 1.00 & 0.563 & 1.00 & 0.568 & 0.988 & 0.567 & 1.00 & 0.608 & 1.00 & 0.591 \\ \hline
256 & 1.00 & 0.660 & 1.00 & 0.688 & 1.00 & 0.694 & 0.996 & 0.623 & 1.00 & 0.658 & 1.00 & 0.652 \\ \hline
1024 & 0.971 & 0.744 & 1.00 & 0.784 & 1.00 & 0.788 & 0.955 & 0.663 & 1.00 & 0.709 & 1.00 & 0.722 \\ \hline
4096 & 0.947 & 0.807 & 0.99 & 0.836 & 1.00 & 0.829 & 0.900 & 0.806 & 0.995 & 0.821 & 1.00 & 0.826 \\ \hline
8192 & 0.899 & 0.838 & 0.986 & 0.856 & 0.99 & 0.842 & 0.864 & 0.830 & 0.984 & 0.837 & 0.99 & 0.858 \\ \hline
16384 & 0.88 & 0.860 & 0.975 & 0.862 & 0.99 & 0.876 & 0.856 & 0.843 & 0.971 & 0.860 & 0.97 & 0.870 \\ \hline
\multicolumn{1}{l}{} & \multicolumn{6}{c}{\textbf{Homogeneous Classification (Adalm Pluto SDR)}} & \multicolumn{6}{c}{\textbf{Heterogeneous Classification}} \\ \hline
\multicolumn{1}{l}{} & \multicolumn{2}{c}{\textbf{Simple CNN}} & \multicolumn{2}{c}{\textbf{Medium CNN}} & \multicolumn{2}{c}{\textbf{Complex CNN}} & \multicolumn{2}{c}{\textbf{Simple CNN}} & \multicolumn{2}{c}{\textbf{Medium CNN}} & \multicolumn{2}{c}{\textbf{Complex CNN}} \\ \hline
\textbf{\begin{tabular}[c]{@{}c@{}}Dataset\\ Size\end{tabular}} & \textbf{\begin{tabular}[c]{@{}c@{}}Training\\ Acc\end{tabular}} & \textbf{\begin{tabular}[c]{@{}c@{}}Test\\ Acc\end{tabular}} & \textbf{\begin{tabular}[c]{@{}c@{}}Training\\ Acc\end{tabular}} & \textbf{\begin{tabular}[c]{@{}c@{}}Test\\ Acc\end{tabular}} & \textbf{\begin{tabular}[c]{@{}c@{}}Training\\ Acc\end{tabular}} & \textbf{\begin{tabular}[c]{@{}c@{}}Test\\ Acc\end{tabular}} & \textbf{\begin{tabular}[c]{@{}c@{}}Training\\ Acc\end{tabular}} & \textbf{\begin{tabular}[c]{@{}c@{}}Test\\ Acc\end{tabular}} & \textbf{\begin{tabular}[c]{@{}c@{}}Training\\ Acc\end{tabular}} & \textbf{\begin{tabular}[c]{@{}c@{}}Test\\ Acc\end{tabular}} & \textbf{\begin{tabular}[c]{@{}c@{}}Training\\ Acc\end{tabular}} & \textbf{\begin{tabular}[c]{@{}c@{}}Test\\ Acc\end{tabular}} \\ \hline
16 & 0.859 & 0.369 & 1.00 & 0.437 & 1.00 & 0.407 & 0.96 & 0.762 & 1.00 & 0.842 & 1.00 & 0.816 \\ \hline
64 & 0.988 & 0.512 & 1.00 & 0.548 & 1.00 & 0.450 & 1.00 & 0.836 & 1.00 & 0.844 & 1.00 & 0.841 \\ \hline
256 & 0.939 & 0.641 & 1.00 & 0.633 & 1.00 & 0.579 & 0.997 & 0.883 & 1.00 & 0.885 & 1.00 & 0.895 \\ \hline
1024 & 0.862 & 0.700 & 0.993 & 0.689 & 1.00 & 0.673 & 0.99 & 0.913 & 1.00 & 0.920 & 1.00 & 0.919 \\ \hline
4096 & 0.820 & 0.718 & 0.959 & 0.699 & 0.99 & 0.706 & 0.97 & 0.937 & 0.99 & 0.954 & 1.00 & 0.952 \\ \hline
8192 & 0.775 & 0.736 & 0.901 & 0.738 & 0.96 & 0.718 & 0.98 & 0.955 & 0.986 & 0.958 & 1.00 & 0.968 \\ \hline
16384 & 0.772 & 0.747 & 0.865 & 0.753 & 0.85 & 0.754 & 0.98 & 0.97 & 0.975 & 0.970 & 1.00 & 0.972 \\ \hline
\end{tabular}
\label{overfit}
\vspace{-0.2in}
\end{table*}

\subsection{CNN Architecture for Interference Classification}
In this section, we describe the architecture of the CNN model and the training parameters used for the interference classification task. The CNN model is a pre-trained \textit{ResNet18} model described in \cite{he2016deep}. The architecture is summarized in Table~\ref{table:interf_class}. The model consists of many convolutional layers, two fully-connected layers, and one output layer, in that order. Leaky ReLU ($\text{Leaky ReLU}(x)=\max\{\alpha x, x\}$, where $\alpha\in(0,1)$ is a preset parameter). All convolutional layers and fully linked layers are subjected to an activation function with  $\alpha=0.2$. The resulting (output) layer then has the softmax function applied to it. All convolutional layers are subject to batch normalization \cite{ioffe2015batch}, however the output layer and fully-connected layers are excluded. Additionally, we use $\text{stride}=2$ in the convolutional layers rather than $2\times 2$ pooling layers for down-sampling. Such changes enhance performance and lower the variance of the results across various training epochs. x is normalized as $x'=x/x_{\max}$, where $x_{\max}$ is the largest input value allowed in $x$. We utilize the Adam optimizer with the suggested default values in \cite{kingma2014adam}. The learning rate is $1\times 10^{-4}$, and the batch size is $64$.

\section{Experimental Setup}
In this section, we describe the datasets used for the investigation of dataset size, model complexity and level of classification. It is worth noting that we only used publicly available datasets for interference classification, while we used our own testbed to generate datasets for the model complexity studies presented in this paper. This is because there are some limitation with using public dataset, for instance, we were not able to control the types of transmitters used to generate the interference. Also, the channel used for transmission cannot be changed since we are using the dataset as is. 

\subsection{Dataset Generated For Model Complexity}
The dataset generation process used for the assessment of model complexity is similar to the one used in \cite{oyedare2019estimating}. The only difference is that more training data was added when compared to the work in \cite{oyedare2019estimating}. We define the process of creating a baseband waveform, transmitting it over a channel, and receiving it as the transmitter-receiver chain. It is important to note that the details of the transmitter-receiver chain can be found in \cite{oyedare2019estimating}. 

\subsubsection{Baseband Waveform Generation}
The hardware devices transmit OFDM packets created using GNU Radio Companion signal processing blocks. This is done by generating a stream of bits (0s and 1s) from a random source. We used an OFDM waveform that receives 10,000 data bits from a random source, which are mapped to the OFDM waveform using a QPSK modulation scheme with an FFT length of 512 and occupying the center 200 subcarriers with a cyclic prefix of 128.


\subsubsection{RF Transmission}
During transmission, GNU Radio and the transmitter hardware are connected through the USRP hardware driver (UHD) hardware support package. All USRP devices are managed and communicated with using a library called UHD. This is done using a GNU Radio block that takes as inputs the sampling rate, RF bandwidth, buffer size, center frequency, physical address of the device, and attenuation. The OFDM waveform is then up-converted to 2.45 GHz as the transmission center frequency and transmitted through the hardware's antenna. The sets of transmitters include four USRP 2921s, four USRP B200s, and four Adalm Pluto software defined radios (SDRs).

\subsubsection{RF Reception}
\label{reception}
On the receiver side, the signal is received when the antenna and the transmitter are on the same channel. We assume that the receiver knows the center frequency and bandwidth of the transmitter and corrects for frequency offset at the receiver. The transmitted signal's center frequency and sample rate are stored in the UHD USRP Source block, which is used by a computer running GNU Radio Companion to down-convert the signal to baseband frequency. After being delivered to a low noise amplifier, the signal is separated into in-phase and quadrature components at baseband. It is then low-pass filtered and transferred to an analog-to-digital converter (ADC). When the ADC process is completed, the digital samples are clocked into an FPGA. After being digitally down-converted using precision frequency tuning, a series of filters are used to decimate the FPGA image. The raw samples are then transmitted to a host computer through the host interface (using an Ethernet gigabit link, in this case) following decimation, made possible by the UHD. The complex samples are collected by the host computer using GNU Radio, and the IQ data is saved to a file and transferred to MATLAB for further processing. 

\subsubsection{Continuous Wavelet Transform (CWT) Signal Preprocessing}
The features in the received signal that can be employed in the classification process are highlighted by the CWT. For CWT, a $M\times N$ matrix of coefficients is generated from $N$ samples for a $N$-length signal, where $M$ represents the number of scales \cite{matlab}. The scales are determined automatically using the wavelength's energy distribution in frequency and time \cite{matlab}. 

The CWT MATLAB function is used to pre-process the signals in order to enhance key transmitter properties in the received signal. After the signal has been transformed, the resultant 2D matrix of size $M \times N$ for each sample is stacked together and sent to the CNN classifier as a three dimensional matrix. 
The dimension of the output of the CWT process for this project is a $38\times100$ matrix of coefficient whereas the input to the CWT is $2\times128$.

\subsection{Datasets for Interference Classification}
\subsubsection{Radio Frequency Interference Dataset}
In this work, we use publicly available wireless interference datasets to evaluate our approach. The RFI dataset used in this study was created by the authors of \cite{ujan2020efficient}. They created the dataset by combining a signal of interest (SOI) with three types of jammers (continuous-wave interference (CWI), multi-continuous-wave interference (MCWI), and chirp interference (CI)) at different signal-to-noise ratios (SNRs). 

\subsubsection{CRAWDAD Dataset}
The second dataset used in this paper was obtained from the Community Resource for Archiving Wireless Data at Dartmouth (CRAWDAD) website. 
This dataset, provided by Schmidt et al. \cite{schmidt2017wireless}, includes traces of IEEE 802.11b/g, IEEE 802.15.4, and Bluetooth packet transmissions with different SNRs in the baseband \cite{owl-interference-20190212}, as well as frequency offsets in the baseband \cite{grunau2018multi}. There are a total of 15 distinct classes, 10 of which are IEEE 802.15.1 devices, 3 of which are IEEE 802.11 devices, and the remaining 2 are IEEE 802.15.4 devices. 

\begin{figure*}[!htb]
\begin{center}
\subfloat[Confusion matrix for the \textit{RFI} Dataset.(\textit{Classification Accuracy: $97.8\%$})] {\label{fig:rfi}\includegraphics[width=0.45\textwidth]{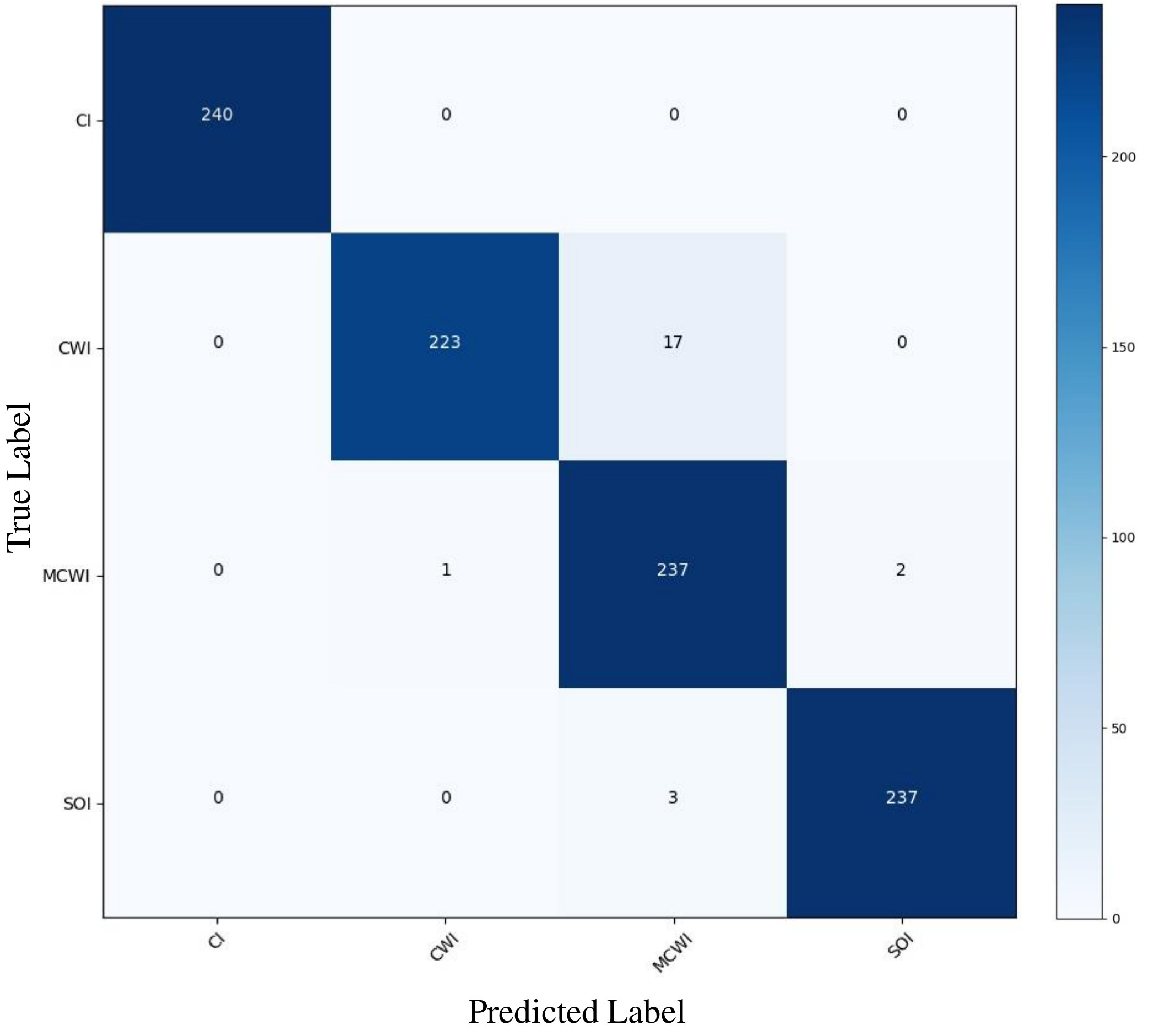}}\hspace{5mm}
\subfloat[Confusion Matrix for the \textit{CRAWDAD} Dataset (\textit{Classification Accuracy: $80\%$})]{\label{fig:crawdad}\includegraphics[width=0.45\textwidth]{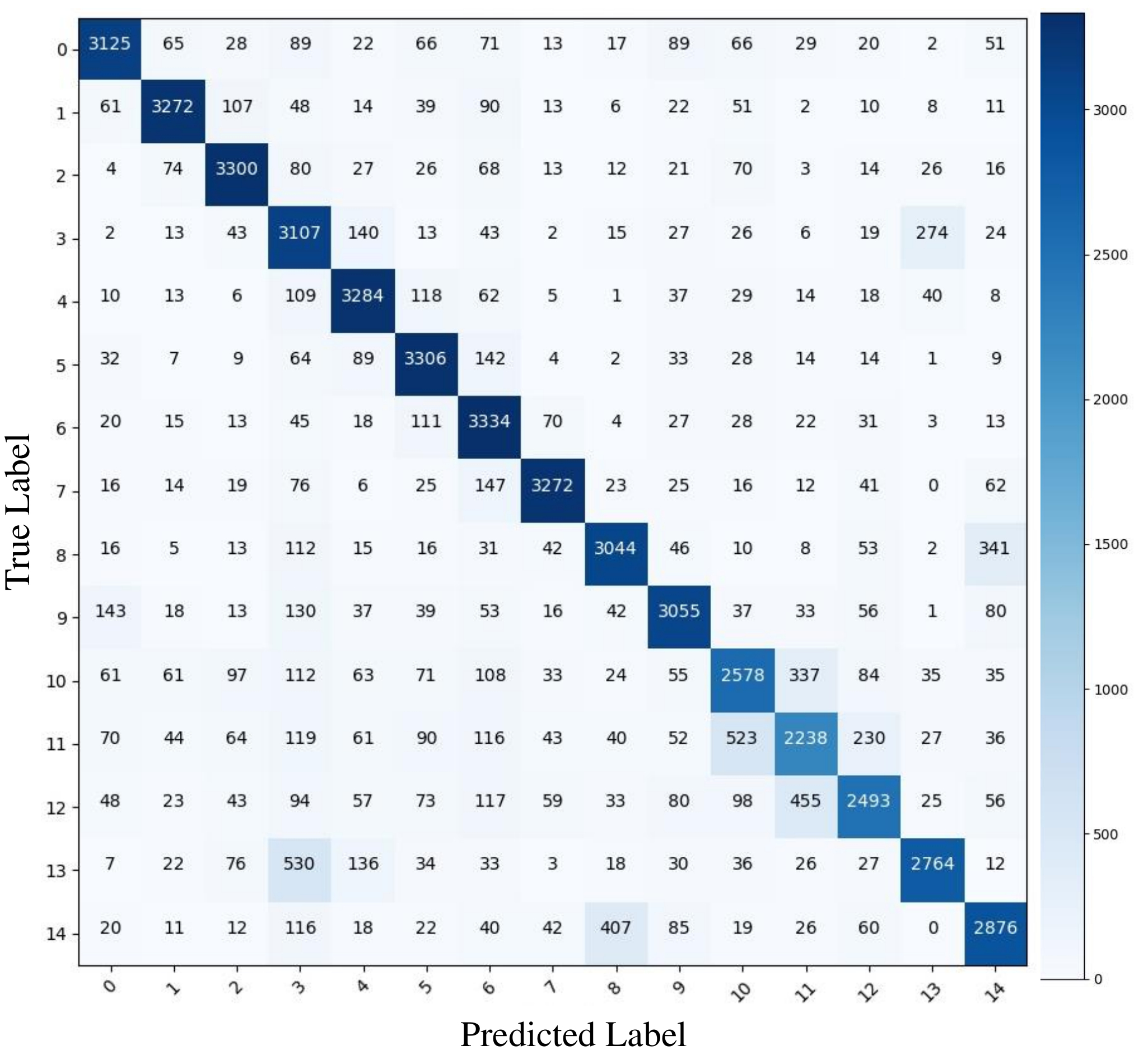}}
\caption{Confusion Matrix for CNN classification of the RFI and CRAWDAD Datasets.}
\label{fig:confusion}
\end{center}
\vspace{-0.15in}
\end{figure*}

\section{Results}
\label{sec:results}
In this section we discuss the results from the model complexity and interference classification studies. Model complexity studies are important since they help to understand the relationship between classification performance and the number of parameters used by a deep learning algorithm (CNN used in this paper). It is important to note that generating our own dataset to study model complexity helps us to vary different conditions in the data generating process which can not be done with publicly available datasets.

\subsection{Model Complexity}
\label{model_comp}
This section examines the performance of the various CNN models across different device categories. When the test accuracy is significantly lower than the training accuracy, it suggests that the algorithm performs well on the training set but poorly on the test set, indicating overfitting. In order for a deep learning algorithm to perform well on new, unseen data, it is important for the training and test accuracies to be similar. This indicates that the algorithm has learned to generalize well.

Table \ref{overfit} compares the training accuracy to the test accuracy for all the classification tasks. At smaller dataset sizes, all the models tend to overfit significantly. However, as the dataset size increases, the algorithms tend to overcome overfitting issues. For most of the classification categories, the overfitting problem is significantly reduced, with the difference between the training and test accuracy being within $5\%$. Overfitting often occurs when the CNN is complex, with multiple layers and many nodes in the hidden layer. To prevent overfitting, regularization techniques such as dropout \cite{xie2016disturblabel} and early stopping \cite{ying2019overview} can be used.

\begin{figure*}[!htb]
\begin{center}
\subfloat[USRP 2921] {\label{fig:2921}\includegraphics[width=0.24\textwidth]{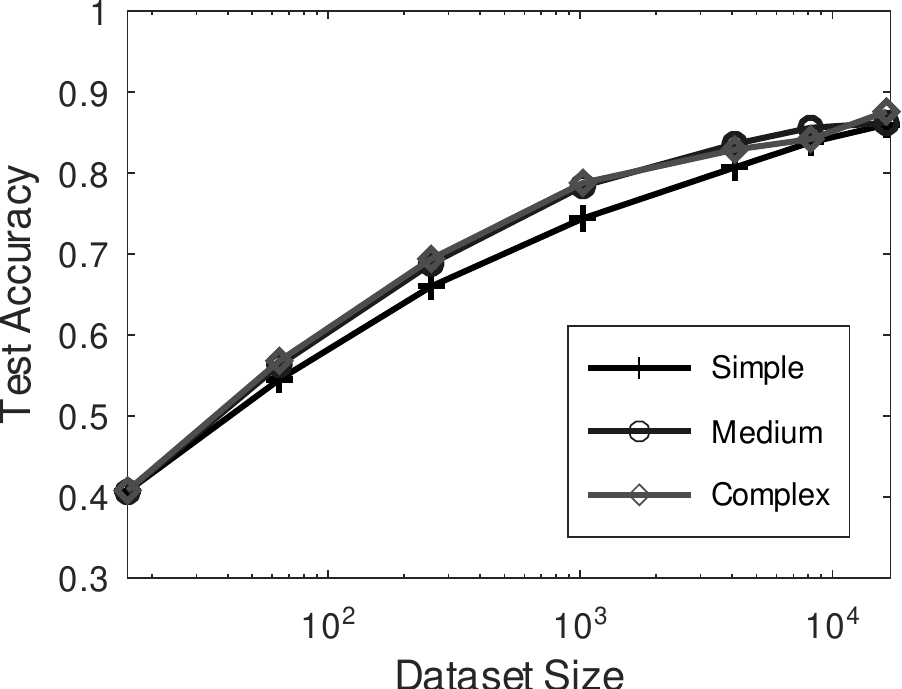}}\hspace{1mm}
\subfloat[USRP B200]{\label{fig:b200}\includegraphics[width=0.24\textwidth]{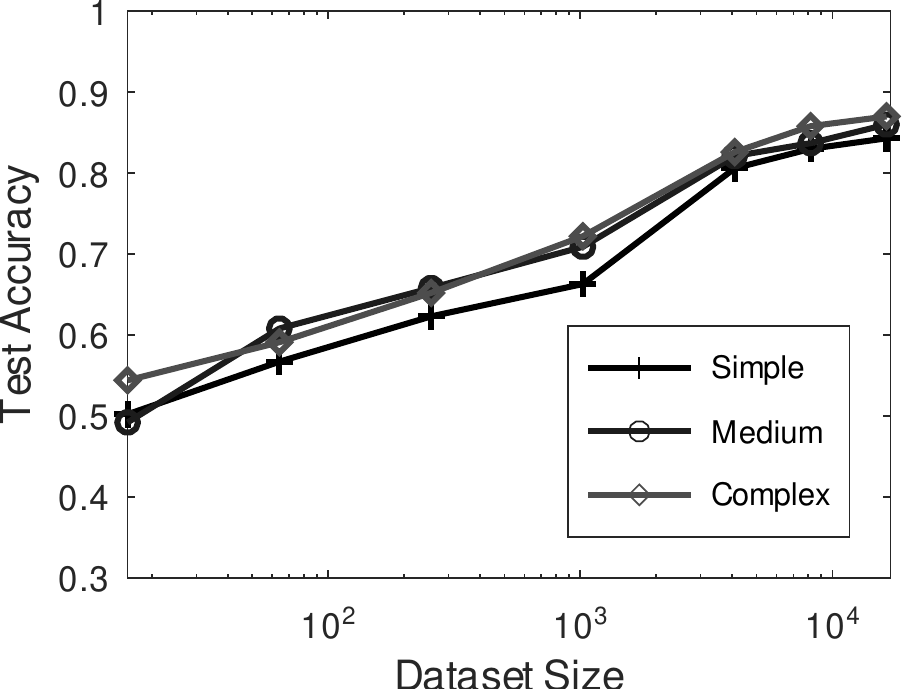}}\hspace{1mm}
\subfloat[Pluto SDR]{\label{fig:pluto}\includegraphics[width=0.24\textwidth]{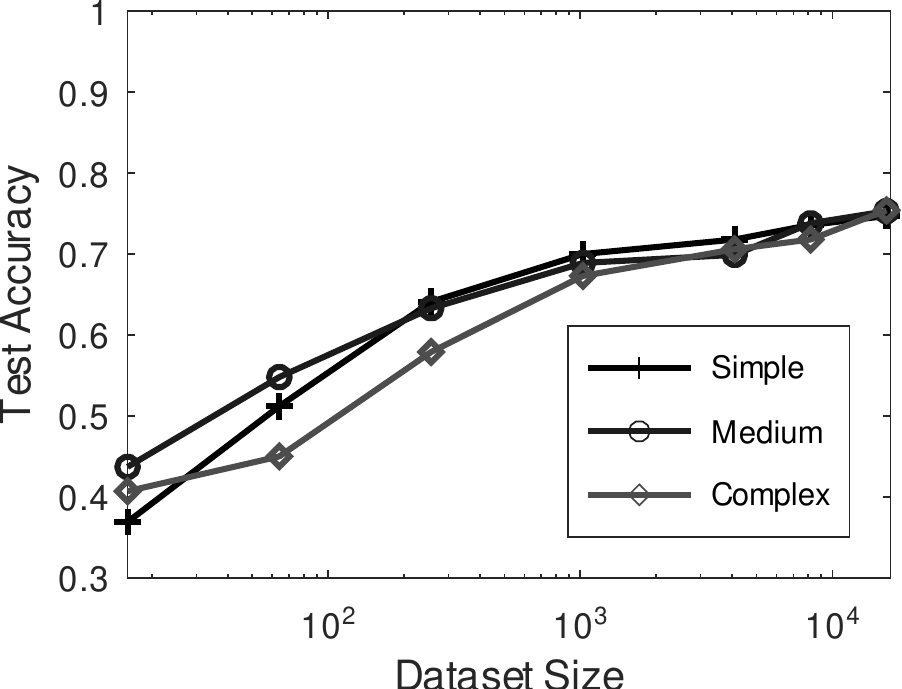}}\hspace{1mm}
\subfloat[Heterogeneous]{\label{fig:heterogeneous}\includegraphics[width=0.24\textwidth]{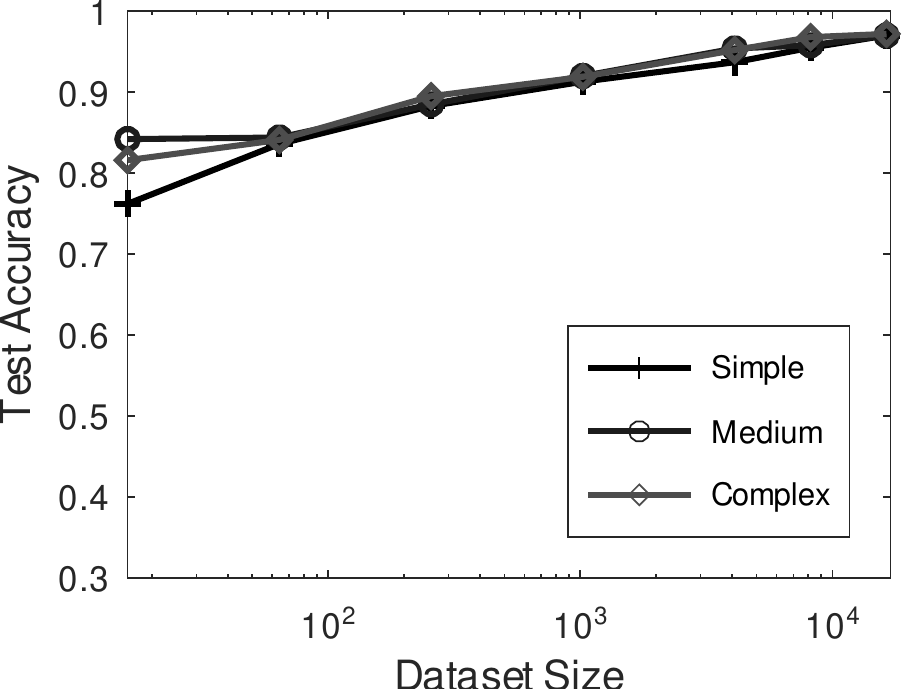}}\hspace{1mm}
\caption{Comparison of Test Accuracy for Homogeneous and Heterogeneous Transmitter Classes}
\label{fig:comparison}
\end{center}
\vspace{-0.2in}
\end{figure*}

Figure \ref{fig:comparison} demonstrates that, for most transmitters, the simple CNN performs similarly to the medium and complex CNNs, despite having significantly fewer parameters. In fact, Figure \ref{fig:comparison}d shows that the performance of the simple CNN is comparable to that of the complex CNNs for heterogeneous transmitters after a dataset size of 100. While more complex or sophisticated CNN models may offer some benefits, there must be a balance between the network's generalization ability and its complexity. \emph{These findings suggest that, it is important for the deep learning model to be both simple and robust, this is especially true for resource-constrained applications}.

\subsection{Interference Classification}
Accurate classification of interference sources is crucial for interference suppression or mitigation. Using a pre-trained ResNet18 model, we achieved a $97.8\%$ accuracy on the RFI dataset, as shown in Table \ref{table:interf_class}. Figure \ref{fig:rfi} shows the confusion matrix for the interference classes, indicating excellent performance on this comparatively easy classification task. For the CRAWDAD dataset, the classification performance was about $80\%$, as shown in the confusion matrix in Figure \ref{fig:confusion}b, after 25 epochs.

As discussed in Section \ref{model_comp}, while more complex CNN models tend to perform better in classification tasks, they are typically prone to overfitting. Overfitting occurs when a deep learning model memorizes the training dataset, leading to high training accuracy but low test or validation accuracy. This can be seen in the results for the Pluto SDR in both Table \ref{overfit} and Figure \ref{fig:comparison}c. These results further emphasize the importance of our findings from the model complexity study discussed in Section \ref{model_comp}. As previously mentioned, a less complex model can often perform as well as a more complex one without the risk of overfitting. Therefore, it is important to use models that are neither too complex nor too simple. The right level of complexity can be determined by using just enough 2D convolution layers and filters to achieve good performance, starting with a simpler model and gradually increasing complexity as needed, this helps to prevent both under-fitting and overfitting issues. 

\section{Conclusion}
\label{sec:conclusion}
In conclusion, this paper has thoroughly examined the use of CNN for interference classification. Our results demonstrate that the CNN model is capable of accurately classifying different interference sources, as shown in the two datasets we used. Our study found that, while medium and complex CNN classifiers performed slightly better than the simple classifier, the difference in performance was not significant. This is an important finding since resource-constrained devices can easily leverage the simpler models. When designing deep learning models, the goal is to create models that can be applied to new data. Simple models are more likely to be able to do this because they are less prone to overfitting, which is a common issue with complex models. Different classification tasks, such as homogeneous and heterogeneous classification, may require different amounts of data and different levels of model complexity. In our study, the training dataset size, number of nodes in the hidden layer has had the greatest impact on CNN model performance. While the size of the dataset is important, we also note that the quality of the training dataset can also significantly impact the performance of CNN classifiers.


\section*{Acknowledgment}
This work was supported in part by the U.S. Air Force Research Laboratory (AFRL) under Grant FA8750-20-2-0504, in part by the Lockheed Martin Corporation under Grant M16-005-RPP010, and in part by the National Science Foundation under Grant CNS-1564148.

\Urlmuskip=0mu plus 1mu\relax
\bibliographystyle{ieeetr}
\bibliography{ref}

\end{document}